\documentclass[showpacs,aps,10pt,twocolumn,groupedaddress,prl]{revtex4-1}
\usepackage{colordvi,epsfig,color,amsmath,amssymb}
\usepackage{color}
\usepackage{hyperref}
\usepackage{ulem}

\begin{document}

\def\be{\begin{equation}}
\def\ee{\end{equation}}
\def\bee{\begin{eqnarray}}
\def\eee{\end{eqnarray}}
\def\kb{k_{\rm B}}
\def\halb{\mbox{$\frac{1}{2}$}}
\def\with{\quad\mbox{with}\quad}
\def\und{\quad\mbox{and}\quad}
\newcommand{\bbbone}{{\mathchoice {\rm 1\mskip -4mu l}{\rm 1\mskip -4mu l}{\rm
1\mskip -4.5mu l}{\rm 1\mskip -5mu l}}}

\definecolor{darkgreen}{rgb}{0,0.5,0}
\definecolor{purple}{rgb}{0.35,0,0.35}
\definecolor{orange}{rgb}{1,0.5,0}
\definecolor{darkred}{rgb}{.7,0,0}
\definecolor{darkblue}{rgb}{0,0,.3}
\definecolor{grey}{rgb}{.6,.6,.6}
\definecolor{dimgreen}{rgb}{0.2,0.6,0.1}

\renewcommand{\emph}[1]{\textit{#1}}

\newcommand{\comment}[1]{{\color{orange}{\textbf{[Comment: #1]}}}}
\newcommand{\todo}[1]{{\color{red}{\textbf{[ToDo: #1]}}}}
\newcommand{\scrap}[1]{{\color{red}{\sout{#1}}}}
\newcommand{\SL}[1]{{\color{blue}{#1}}}

\title{Nonequilibrium Landau-Zener-St\"uckelberg spectroscopy in a double quantum dot}

\author{P. Nalbach$^1$, J. Kn\"orzer$^1$, and S. Ludwig$^2$}
\affiliation{$^1$I.\ Institut f\"ur Theoretische Physik,  Universit\"at Hamburg,
Jungiusstra{\ss}e 9, 20355 Hamburg, Germany \\
$^2$ Center for NanoScience and Fakult\"at f\"ur Physik, Ludwig-Maximilians-Universit\"at
M\"unchen, Geschwister-Scholl-Platz 1, D-80539 M\"unchen, Germany}

\date{\today}

\begin{abstract}
We study theoretically nonequilibrium Landau-Zener-St\"uckelberg (LZS) dynamics in
a driven double quantum dot (DQD) including dephasing and, importantly, energy
relaxation due to environmental fluctuations. We derive effective nonequilibrium
Bloch equations. These allow us to identify clear signatures for LZS oscilations
observed but not recognized as such in experiments 
[Petersson et al., Phys. Rev. Lett. 105, 246804, 2010] and to identify the full
environmental fluctuation spectra acting on a DQD given experimental data as in
[Petersson et al., Phys. Rev. Lett. 105, 246804, 2010]. Herein we
find that super-Ohmic fluctuations, typically due to phonons, are the main relaxation
channel for a detuned DQD whereas Ohmic fluctuations dominate at zero detuning.
\end{abstract}


\maketitle

Quantum electronic devices, as qubits realized by double quantum dots (DQD),
require coherence times which exceed their quantum
operation time during which the DQD is typically strongly driven by external
voltage pulses. Tremendous research efforts studied semiconductor based devices
to achieve coherent quantum control
\cite{Pet10,Fuj06,Hans07,Pio08,Fuj09,Gaud11,Bluhm11}.
Many fluctuation sources of the noisy solid state environment, which act on the
electron in the DQD and thus destroy coherence, were revealed but a comphrehensive
picture is elusive. Furthermore, driving by voltage pulses causes an intrinsic
nonequilibrium situation in which relaxation competes with driving
\cite{Nal09a,Nal10c,Har10} which renders a theoretical description of the dissipative
nonequilibrium dynamics highly nontrivial.

Here, we theoretically study the dissipative nonequilibrium dynamics of a single
electron charge qubit defined in a DQD embedded in a noisy solid state environment
driven by voltage pulses.
While DQD charge qubits have relatively short coherence times, this disadvantage is 
compensated by the possibility of fast quantum operations. We model the DQD and its 
dissipation as a quantum
two-level system in an open quantum system approach \cite{WeissBuch}. We determine
the dissipative nonequilibrium real time dynamics (initialized by applying voltage
pulses) by deriving effective nonequilibrium Bloch equations (NBEs). These allow
fast numerical treatment in contrast to numerical exact methods \cite{Nal09a,Nal10c}
and thus allow a comphrehensive analysis of recent experiments by Petersson et al.\
\cite{Pet10} and Dovzhenko et al.\ \cite{Dov11}. In these experiments quantum
control of a single electron was achieved by means of applying ultra short voltage
pulses to control gates of the laterally defined DQD. In these time ensemble
measurements the DQD was cycled (with 40\,MHz repetition rate) between two
different ground state configurations while its average charge occupation was
continuously detected via the electric current through a capacitively coupled quantum
point contact (QPC). The applied voltage pulses generate Landau-Zener-St\"uckelberg (LZS)
dynamics \cite{LZallg} and we can identify so far unexplained
features in the experimental data as signatures of coherent LZS oscillations.

In the experimental ensemble measurements the dephasing time $T_2^\star$ due to
slow noise \cite{Bluhm11,Lee08} is much shorter than relaxation times $T_1$ and thus
dominates the decoherence times $T_2$ since $T_2^{-1}={T_2^\star}^{-1}+(2T_1)^{-1}$.
Relaxation and dephasing could be caused by thermal phonons, intrinsic
or externally triggered charge noise \cite{Pio05,Taubert2008}, or
detector back-action \cite{Aguado2000,Young2010,Taubert2008,Gasser2009,Schi09,Lu10,Lu12}.
Ref. \cite{Pet10,Dov11} neglected relaxation and employed solely an $1/f$ noise model
\cite{Alt09,Alt10,Pal,Nal10b,Nal12b} for dephasing which allowed them to describe the detuning
dependence of the observed decoherence of the DQD.

The experimentally observed steady state charge occupation of the DQD is, however, heavily
influenced by relaxation \cite{Har10}. Therefore, in this letter, we go beyond this
simple dephasing model and include various environmental fluctuation spectra to
describe dephasing and relaxation. This allows us to simulate the full nonequilibrium
real time dynamics experimentally studied. We find that the observed visibilty reduction is 
caused by relaxation. Moreover, by analyzing the measured LZS dynamics we identify the full environmental
fluctuation spectrum as a sum of three processes. In addition to slow noise, already 
considered in ref.\  \cite{Pet10}, which causes detuning dependent dephasing, super-Ohmic 
fluctuations, as typically originating from phonons, are the main relaxation channel for the 
detuned DQD. Near zero detuning, however, Ohmic fluctuations dominate
relaxation. The latter also limit the experimentally observed maximal decoherence time of $T_2\sim7\,$ns.

\paragraph{Modelling of pulse driven DQD}

\begin{figure}[t]
\epsfig{file=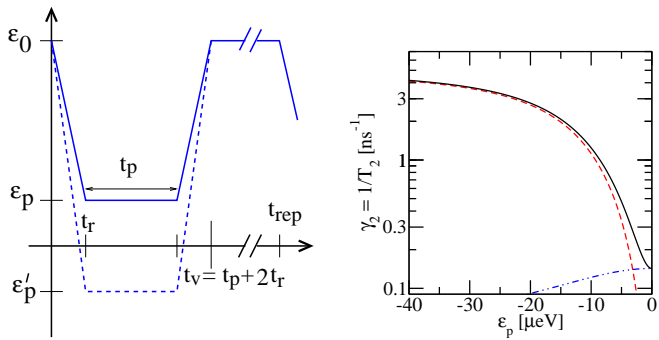,width=4.1cm}\hfill\epsfig{file=Fig1bNalbach.eps,width=4.1cm}
\caption{\label{fig1} (Left): Energy detuning profile for a voltage pulse with rise time
$t_r$, pulse plateau time $t_p$, pulse duration $t_v=t_p+2 t_r$. If cycled the pulse
periodically repeats with repetition time $t_{\rm rep}$. (Right): Decoherence rate versus
detuning $\epsilon_p$. Full theory (black full line), Ohmic dephasing (red dashed line) and
decoherence due to relaxation (blue dot-dashed line).}
\end{figure}

We model the single electron DQD using a two-level system Hamiltonian
\be\label{hamtls} H = \halb\Delta \sigma_x + \halb \epsilon(t)\sigma_z
\ee
where the eigenstates of $\sigma_z$ correspond to the electron in the left / right dot, $\epsilon(t)$
is the level detuning and $\Delta$ is the interdot tunnel splitting \cite{foot1}.
Experimentally, gate voltage pulses are applied to change the level detuning as sketched in
Fig.\ \ref{fig1}(left). Initially the detuning is $\epsilon_0=\epsilon(0)\gg\Delta$ and the
DQD in the according ground state (0,1) [electron is in the right dot]. A voltage pulse then
drives the system within a rise time $t_r$ to a plateau detuning $\epsilon_p$ close to
{\it resonance} ($\epsilon=0$) and keeps it there for a plateau time $t_p$. Then the DQD is
driven back within $t_r$ to the initial detuning and the probability $P_{(1,0)}$ of occupation
of the excited state (1,0) [electron in the left dot] is studied as a function of the pulse
duration $t_v=2t_r+t_p$. For a quantum mechanical two-level system without dissipation we
expect coherent oscillations in the voltage pulse time $t_v$.

A voltage change on a single gate \cite{Pet10,Dov11} not only
affects the level detuning but also causes a common energy shift of both
states $(1,0)$ and $(0,1)$ which is not relevant in the
following and thus neglected. The sharp corners of $\epsilon(t)$ in
Fig.\ \ref{fig1}(left) are smoother in reality and might even contain
oscillatory features (due to a finite band-width transfer function).
At voltage pulse times $t_v\gtrsim 2t_r$ deviations between
linear voltage ramps and more accurate descriptions are negligible.

\begin{figure}[t]
\epsfig{file=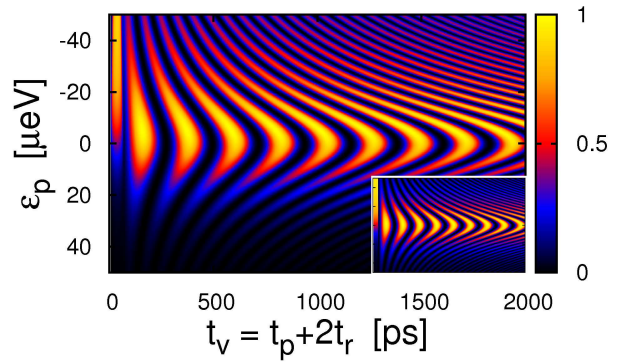,width=8cm}
\caption{\label{fig2} $P_{(1,0)}(t_v)$ for the fully
coherent case ($T_1=T_2=T_2^\star=\infty$), right after a single voltage pulse,
vs. level detuning $\epsilon_p$ and pulse duration $t_v=t_p+2t_r$ for
$\epsilon_0=200\mu\text{eV}$ ($\epsilon_0=2000\mu\text{eV}$) for the main figure (inset).}
\end{figure}

Solving the quantum dynamics of the DQD with a single applied voltage
pulse (details given below) results in the probability $P_{(1,0)}(t=t_v)$
after the voltage pulse, plotted in Fig.\ \ref{fig2} as a function of
$t_v$ and plateau detuning $\epsilon_p$. In order to model the experiments
\cite{Pet10,Dov11} we use $\Delta/h=4.5\,$GHz,
$\epsilon_0=200\mu\text{eV}\simeq 11\Delta$ and $t_r=35\,$ps which
corresponds to the fastest experimental achievable rise time \cite{foot3}.
The sweep speed $v_p=|\epsilon_p-\epsilon_0|/t_r$ can be approximated by
$v_0=|\epsilon_0|/t_r$ for $\epsilon_0\gg\epsilon_p$ [slope in Fig.\
\ref{fig1}(left)] which provides an indication on the overall adiabaticity
of the dynamics \cite{LZallg}.
For $v_0=200\mu\text{eV}/35\text{ps}\simeq 11\Delta^2/\hbar$, Fig.\ \ref{fig2}
shows coherent oscillations between states (0,1) and (1,0) with frequency
$E/h$ and eigenenergy $E=\sqrt{\Delta^2+\epsilon_p^2}$ of Eq.\ (\ref{hamtls}).
At zero detuning $P_{(1,0)}(t_v)$ oscillates between $0$ and $1$. With
increased $\epsilon_p$ the oscillation frequency grows while its amplitude
gradually decreases. Fig.\ \ref{fig2} shows a clear asymmetry: the visibility
is larger for $\epsilon_p<0$ compared to $\epsilon_p>0$. Remarkably,
this asymmetry was observed experimentally, too (but not explained) \cite{Pet10}.
The asymmetry decreases with increasing sweep speed as highlighted by the inset
of Fig.\ \ref{fig2} which shows an almost symmetric $P_{(1,0)}(t_v)$ for
$\epsilon_0=2000~\mu$eV leading to $v_0\gtrsim 110\Delta^2/\hbar$.
We conclude that the asymmetry is solely an effect of adiabaticity when the
DQD is driven through the avoided crossing at zero detuning.
For $\epsilon_p\lesssim0$, the quantum system accumulates, during the voltage
pulse, not only phase due to the coherent oscillation of the electron between
the two dots but also due to the superposition state occupied between two
successive Landau-Zener transitions \cite{LZallg}.
As such, the asymmetry is a clear signature of coherent LZS oscillations.

\paragraph{Periodically Cycled Pulses}

In an ensemble measurement with continuous charge detection, as in Ref.\ \cite{Pet10}
with $t_{\rm rep}=25\,$ns, it is essential to choose $t_\text{rep}\gg t_v$ to ensure
readout of the charge occupation \emph{after} application of the pulses (of duration
$t_v\lesssim T_2$).
An interpretation in terms of an ensemble measurement, which simply averages over many
independent shots, further requires $t_\text{rep}\gg T_2$ and, interestingly,
$t_\text{rep}\sim T_1$, where $T_1$ is the (thermal) energy relaxation time which
depends on the detuning: for $t_\text{rep}\gg T_1$ initialization into configuration
(1,0) is guaranteed, but the visibility of the continuous measurement is close to zero
as mostly (1,0) is occupied; for $t_\text{rep}\ll T_1$ initialization into (1,0)
independently of $P_{(1,0)}$ --right after the pulse-- is impossible. No matter of
the choice of pulse sequence, continuously cycled pulses will result in a steady
state which, in principle, contains the information of dephasing and energy relaxation
times \cite{Har10}.

In order to model the experimental repeated pulse train sequence we repeat our
simulation after the first cycle (up to the repetition time $t_{\rm rep}$) with
the final statistical operator of the previous cycle as initial state. This procedure
is repeated until the population $\bar{P}_{(1,0)}$ [measured as average over a full
cycle as plotted in Fig.\ \ref{fig1}(left)] changes by less than $0.005$. This
$\bar{P}_{(1,0)}$ approximates the experimentally observed steady state
population.

\paragraph{Driven dissipative dynamics}
%
\begin{figure}[t!]
\epsfig{file=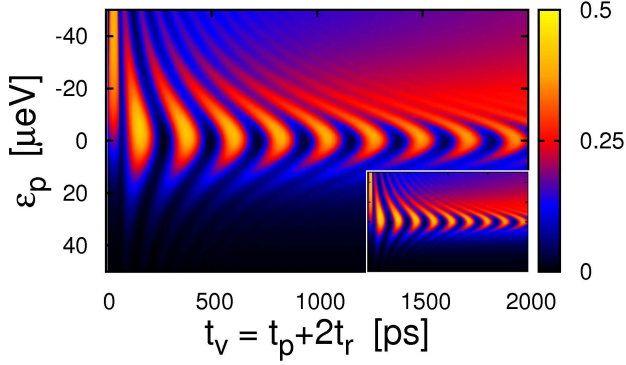,width=8.3cm}
\caption{\label{fig4} (Main): $\bar{P}_{(1,0)}$ [(Inset): $P_{(1,0)}(t_v)$ with color scale from 0 to 1] 
versus detuning $\epsilon_p$
and voltage pulse time $t_v=t_p+2t_r$ for a driven DQD with $\epsilon_0=200\mu\text{eV}$
including relaxation and dephasing. 
}
\end{figure}

Including dissipation into the driven dynamics of the DQD causes a competition
between driving and relaxation \cite{Nal09a,Nal10c} which renders all but expansive
numerical treatments inadequate. Following
the standard approach within open quantum dynamics \cite{WeissBuch} we couple
the driven two-level Hamiltonian (\ref{hamtls}) of the DQD to environmental
fluctuations described by harmonic oscillators. This results in
\be\label{hamtot} H_{\rm tot} = H(t) + \frac{\sigma_z}{2}\sum_k\lambda_k (b_k+b_k^\dagger)
\,+\sum_k \omega_k b_k^\dagger b_k
\ee
with bosonic annihilation/creation operators $b_k/b_k^\dagger$. The spectrum
$G(\omega)=\sum_k\lambda_k^2\delta(\omega-\omega_k)$ is typically a smooth
function \cite{WeissBuch} at the energies of interest, i.\,e.\
$G(\omega)=2\alpha\omega^s \exp(-\omega/\omega_c)$
with spectral exponent $s$, cut-off frequency $\omega_c$ and the coupling strength $\alpha$.
Typically\cite{Pet10}, $\Delta\gg\gamma_2\gg\gamma_1$ with the relaxation rate
$\gamma_1=1/T_1$ and the decoherence rate $\gamma_2=1/T_2$, which justifies a weak
DQD-environment coupling approach.
Due to the time dependence of $H(t)$ the weak coupling Born-Markov approximation, however,
fails. Assuming that the change of energy during the memory time in the environment
is small, allows an additional adiabatic rate approximation. For this, we switch to the time dependent
basis where $H(t)=\halb E(t)\tau_x$ is diagonal with Pauli matrizes $\tau_j$ describing
the eigenstates with energy difference $E(t)=\sqrt{\Delta^2+\epsilon(t)^2}$. Employing a
lowest order Born approximation for the memory kernel in the system-bath coupling \cite{Nal02,Nal10a},
one obtains for the time evolution of the components of the statistical operator,
$\rho(t)=\halb(\bbbone+\sum_jr_j\tau_j)$, nonequilibrium Bloch equations
\be\label{blocheq} \begin{array}{lcrrl}
     \partial_t r_x(t) & = &    \phi'(t) r_z(t) &                       & \hspace*{-1.5cm}-\gamma_1(t) [r_x(t)-r_x^{eq}(t)]  \\
     \partial_t r_y(t) & = &  -\gamma_2(t) r_y(t) &         -E(t) r_z(t)                                                    \\
     \partial_t r_z(t) & = &         E(t) r_y(t) &  -\gamma_2(t) r_z(t)  &                   -\phi'(t) r_x(t)
    \end{array}
\ee
with time dependent momentary equilibrium $r_x^{eq}(t)=\tanh[\beta \halb E(t)]$ and time dependent decay rates
\bee \label{gamrel} \gamma_1(t) &=& \frac{\pi}{2\hbar}\coth[\halb\beta E(t)] u^2(t)G[E(t)]  \\
\label{gamdec}\gamma_2(t) &=& \halb\gamma_1(t) \,+ \Gamma_2(t)
\eee
where $\Gamma_2(t)=1/T_2^\star(t)$ is the dephasing time. Herein, $\phi(t)=\arctan[\epsilon(t)/\Delta]$,
$u(t)=\cos\phi(t)$, and $v(t)=\sin\phi(t)$. The presented nonequilibrium Bloch equations with time-dependent
rates and momentary equilibrium describe adequately dissipative Landau-Zener dynamics, i.e. the regime of
competition between driving and relaxation, for weak DQD-environment coupling. We ensured this by extensive
comparisson with numerical exact results \cite{Nal09a,Nal10c}. Details will be presented elsewhere.

To proceed, we need the spectra of fluctuations acting on the DQD. Estimates can be gained by the
observations for fixed DQD parameters of Petersson et al. \cite{Pet10}
of a relaxation time $T_1(\epsilon_0)=10\,$ns at the initial detuning $\epsilon_0$, a decoherence time of
$T_2\sim7\,$ns at zero detuning and the detuning dependent dephasing times presented in Fig.4c of Ref.\ \cite{Pet10};
all at the temperature $T=80\,$mK.

\paragraph{Relaxation rates}
\begin{figure}[t!]
\epsfig{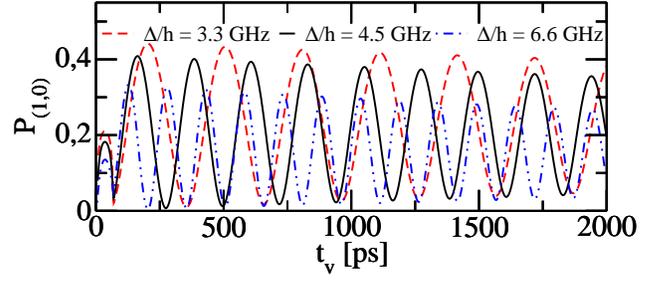}
\caption{\label{fig5} $\bar{P}_{(1,0)}$ for zero detuning versus voltage pulse time $t_v=t_p+2t_r$
for various tunnel couplings.}
\end{figure}

Charge fluctuations and bulk phonons likewise couple to the DQD. In order to include both, we consider
an Ohmic fluctuation spectrum with $s=1$ (charge fluctuations) as well as a super-Ohmic one with $s=3$
(phonons) \cite{WeissBuch}. The rate (\ref{gamrel}) reflects a one-boson process and
accordingly only yields a substantial relaxation if the spectrum at the eigenenergy of the DQD is
finite, i.\,e.\ $E(t)\ll\omega_c$.

At fixed $T$, $\Delta$ and $\epsilon_0$ we can determine the coupling strength $\alpha$ (assuming $E(t)\ll\omega_c$)
using the known energy relaxation time $T_1(\epsilon_0)$,
and estimate $T_1(\epsilon=0)$ using Eq.\ (\ref{gamrel}) and $u(t)=\Delta/E(t)$: For Ohmic fluctuations
$T_1^{s=1}(\epsilon=0)=(\Delta/\sqrt{\Delta^2+\epsilon_0^2})\cdot T_1(\epsilon_0)$ while for super-Ohmic
fluctuations $T_1^{s=3}(\epsilon=0)=(\sqrt{\Delta^2+\epsilon_0^2}/\Delta)\cdot T_1(\epsilon_0)$.
With the reported $T_1(\epsilon_0)=10\,$ns we find $T_1^{s=1}(\epsilon=0)\simeq 900\,$ps assuming
Ohmic fluctuations compared to $T_1^{s=3}(\epsilon=0)\simeq110\,$ns assuming super-Ohmic fluctuations.
According to Eq.\ (\ref{gamdec}) energy relaxation causes an upper bound of the decoherence $T_2\le 2 T_1$.
The decoherence time of $T_2=7\,$ns, observed in the experiment at $\epsilon\simeq0$ \cite{Pet10}, lies
well in between our predictions $T_1^{s=1}\ll 7\,\text{ns}\ll T_1^{s=3}$.
An Ohmic fluctuation spectrum alone --consistent with $T_1(\epsilon_0)=10\,$ns-- would result in
much too fast decoherence at $\epsilon=0$ compared to the experimental results. Hence, super-Ohmic
fluctuations, typically caused by phonons, are the main relaxation mechanism at $\epsilon_0$.

\paragraph{Decoherence and Dephasing rates}

For a super-Ohmic spectrum $\Gamma_2(t)\equiv 0$, and the decoherence rate $\gamma_2$
[see Eq.\ (\ref{gamdec})] would be solely determined by the energy relaxation mechanism.
The spectrum of charge noise, which is well known to cause additional dephasing, strongly
depends on the sample and how it has been treated \cite{Pio05}. Often, charge noise can
be assumed to be slow noise as done by Petersson et al.\ \cite{Pet10}. They use
an $1/f$ dephasing model \cite{Pal} which typically results from background charge
noise \cite{Alt09,Alt10} and might be described in terms of sub-Ohmic noise \cite{Nal12b,Nal10b}
with $s=0$ and $\omega_c\ll\Delta$.
Such \emph{slow} noise is a major dephasing source in realistic devices and causes solely dephasing.
We aim at unraveling the relaxation mechanisms present rather than the dephasing sources. Relaxation
is not influenced by slow noise and thus we describe slow noise here using a simplified Ohmic slow
noise model resulting in $\Gamma_2(t)=(4\pi/\hbar) v^2(t)\alpha_2  \kb T$ characterized by a coupling
strength $\alpha_2$ which we fit to the experiment.
Our Ohmic slow noise model captures two important points, namely, that it does not influence
relaxation, i.e. $\omega_{c2}\ll\Delta$, and that it couples to the position of the charge in
the DQD leading to $\Gamma_2(t)\propto v^2(t)=\epsilon^2(t)/(\Delta^2+\epsilon^2(t))$, which coincides
with the dependence used by Petersson et al.\cite{Pet10} in their $1/f$ noise model.
As typically done, we, herein, neglect fluctuations in $\Delta$.

The red dashed line in Fig.\ \ref{fig1}(right) represents the contribution to decoherence by our slow
Ohmic dephasing noise and thus reflects most of the decoherence (black full line in Fig.\ \ref{fig1}(right)
which reproduce the measured data in  Fig.4c of Ref.\ \cite{Pet10} remarkebly well) except at very small
detunings $\epsilon_p\ll\Delta$.
For $v(t)=\epsilon_p=0$, however, the slow noise dephasing rate $\Gamma_2(t)\equiv 0$
and $\gamma_2(t)=\halb\gamma_1(t)$. Super-Ohmic fluctuations result
in $T_1^{s=3}(\epsilon=0)\simeq110\,$ns, much longer than the zero detuning decoherence
time $T_2=7\,$ns actually measured. To resolve this discrepancy, we add a third noise
source, namely weak Ohmic fluctuations which contribute to relaxation at zero detuning in
addition to the super-Ohmic contribution of phonons. The
contribution of weak Ohmic fluctuations to energy relaxation at large detuning is very small.
The physical origin
of this mechanism could be related to fast local potential fluctuations (e.\,g.\ via voltage
noise on the lead gates).

Our complete model includes three fluctuations spectra
\be\label{spec}
G(\omega)=2\alpha_1\omega^3 e^{-\omega/\omega_{c1}}
+\sum_{j=2}^3 2\alpha_j\omega e^{-\omega/\omega_{cj}}
\ee
with $\omega_{c1},\omega_{c3}\gg\Delta$ and $\omega_{c2}\ll\Delta$. Choosing
$\alpha_1\Delta^2=8.09\cdot 10^{-5}$, $\alpha_2=3.53\cdot 10^{-2}$
and $\alpha_3=2.73\cdot 10^{-3}$ for the coupling strengths reproduces the experimental
observations \cite{Pet10} of $T_1(\epsilon_0)=10\,$ns, $T_2(\epsilon=0)\sim7\,$ns and
the decoherence time versus detuning with measured
data in  Fig.4c of Ref.\ \cite{Pet10} and our predicted
decoherence rate plotted as black full line in Fig.\ \ref{fig1}(right).

\paragraph{Dynamics of the driven dissipative DQD}

Solving the nonequilibrium Bloch equations (\ref{blocheq}) with the fluctuation
spectrum (\ref{spec}) numerically results in Fig.\ \ref{fig4}, the main result,
where we plot the steady state occupation $\bar{P}_{(1,0)}$ versus
voltage pulse time and plateau detuning for a tunnel coupling $\Delta/h=4.5$GHz.
In comparison, the inset shows the instantaneous probability $P_{(1,0)}(t=t_v)$
with the color code stretched by a factor of two, i.\,e.\ {\sl yellow} $=1$. The 
continuous current measurement in the QPC reduces the visibility for the
coherent dynamics by a factor of $2$.
In comparison to the undamped case in Fig.\ \ref{fig2} we observe an overall
reduction in oscillation amplitude. On top, the LZS oscillations smear out
with both, more negative detunings $\epsilon_p$ and
longer voltage pulse times $t_v$: $\bar{P}_{(1,0)}$ is flat in the upper right
corner of Fig.\ \ref{fig4}.
The same behaviour was experimentally observed, but
not explained, by Petersson et al.\ \cite{Pet10}. Our results indicate that
the LZS oscillations smear out as a result of
relaxation but the remaining asymmetry with respect to $\epsilon_p$ is a result
of adiabaticity. Thus, the experimentally observed feature is a clear signature
of LZS oscillations.
%
The overall qualitative agreement between our simulation results and the experimental
data \cite{Pet10} is very good. The quantitative overestimation of the oscillation
amplitude being a factor of $2$ is likely due to uncertainties regarding experimental
details \cite{foot4}.

\paragraph{Interdot tunnel coupling}

Petersson et al. \cite{Pet10} observe an increased (decreased) oscillation
amplitude for a smaller (larger) interdot tunnel
coupling, i.\,e.\ $\Delta/h=3.3\,$GHz (6.6\,GHz). We find
the same tendency. It results from the fact that the relaxation rate [eq.\ (\ref{gamrel})]
increases strongly with the tunnel coupling. Fig.\ \ref{fig5} plots our predictions
for $\bar{P}_{(1,0)}$ as a function of $t_v$ for zero detuning for the three different
tunnel couplings.

\paragraph{Conclusions}

We studied theoretically the dissipative nonequilibrium dynamics of a single
electron DQD driven by voltage pulses. We couple the DQD additionally to 
environmental fluctuations causing
relaxation and dephasing. Extending standard Born-Markov approaches to driven
systems, we derive nonequilibrium Bloch equations exhibiting time dependent
rates and the momentary equilibrium. This approach allows efficient numerical 
simulations of the full nonequilibrium real time dynamics and a comprehensive 
analysis of the experimental data \cite{Pet10}.
We identify an asymmetric occupation of the left dot in respect to the detuning 
between the dots in ref.\ \cite{Pet10} as a clear experimental signature of LZS dynamics. 
A full analysis of the LZS dynamics furthermore allows us to specify the full environmental
fluctuation spectrum acting on the DQD studied by Petersson et al.\ \cite{Pet10}
as sum of three processes. Besides slow noise causing the detuning dependent
strong dephasing, super-Ohmic fluctuations, as typically originating from phonons,
are the main relaxation channel for a detuned DQD. At zero detuning, however,
Ohmic fluctuations, which might be caused by gate voltage noise, dominate relaxation 
and are also the main cause for the decoherence at zero detuning.

\paragraph{Achnowledgements}
We thank M. Thorwart for continuous financial and intellectual support. 
J.K. acknowlegdes support from the DFG via SFB-925. S.L. acknowledges support 
from the DFG via SFB-631, the Cluster of Excellence Nanosystems Initiative 
Munich, and a Heisenberg fellowship.


\begin{thebibliography}{99}
\bibitem{foot1} Notice that our definition $\Delta$ includes a factor of $2$ in comparisson to Ref.\cite{Pet10}.
\bibitem{foot3} Longer rise times up to $100$ps under nonoptimal experimental conditions render the dynamics more adiabatic
and thus would only strengthen our arguments below.
\bibitem{foot4} Better quantitative agreement could be observed using $\epsilon_0\gtrsim\Delta$. Then, relaxation for the 
DQD (detuned and symmetric) could be described by Ohmic fluctuations alone. However, experimentally 
$\epsilon_0\gg\Delta$ \cite{Pet10}.
\end{thebibliography}
\end{document}